\documentclass[11pt,twoside]{article}
\usepackage{./asp2004}
\usepackage{psfig}
\usepackage{epsf}
\usepackage{graphics}
\usepackage{lscape}
\def\edcomment#1{\iffalse\marginpar{\raggedright\sl#1\/}\else\relax\fi}
\reversemarginpar

\begin{document}
\title{Extra-planar Diffuse Hot Gas Around Normal Disk Galaxies}
\author{Q. Daniel Wang}
\affil{University of Massachusetts Amherst, USA}

\begin{abstract}

I review results from {\sl Chandra} observations of nearby normal
edge-on galaxies (Sd to Sa types). These galaxies have a broad
range of star formation rate, but none of them is dominated by a 
nuclear starburst. The galaxies are all in directions of 
low Galactic foreground absorption ($N_{HI} \la 4
\times 10^{20} {\rm~cm^{-2}}$). Extra-Planar diffuse
soft X-ray emission is detected unambiguously
from all the galaxies, except for N4244 (Sd), which is low in both
the stellar mass and the star formation rate. The thermal
nature of the X-ray-emitting gas is well established, although its 
chemical and ionization states remain largely uncertain. 
The X-ray luminosity of the gas is proportional to the star 
formation rate and to the stellar mass of the galaxies. 
But the luminosity accounts for at most a few percent 
of the expected supernova mechanical energy input.
Therefore, there is a ``missing'' energy problem for spiral galaxies.
Much of the energy in late-type spirals may be converted and radiated
in lower energy bands. But early-type ones most likely have outflows,
which are
powered primarily by Type Ia supernovae in galactic bulges. These galactic 
outflows may strongly affect both the dynamics and cooling
of the intergalactic gas accretion, hence the evolution of the galaxies.

\end{abstract}
\thispagestyle{plain}

\section{Introduction}

It has long been theorized that a major, possibly dominant, phase of
the interstellar medium (ISM) is gas at temperature $T \sim 10^6$ K in virtually all galaxies. 
This hot ISM is thought to be created and maintained primarily by 
supernova explosions (SNe; e.g., McKee \& Ostriker 1977). 
If sufficiently energetic, the hot gas is expected to flow 
outward, creating a large-scale gaseous corona (e.g., Spitzer 1956; 
Bregman 1980a, Norman \& Ikeuchi 1989) or
even escaping as a galactic wind (e.g., Bregman 1980b). Therefore, 
the study of extra-planar hot gas is fundamentally important to our
understanding of the mass, energy, and chemical evolution of a disk galaxy such
as our own.

Observationally, although an
understanding of the large-scale properties of the hot ISM in our own 
Galaxy is still elusive, the presence of large amounts of diffuse hot gas
 in nearby edge-on disk galaxies has been established. Much 
of this progress has been made with recent {\sl Chandra} 
observations, which provide high-resolution 
and panoramic views of extra-planar hot gas and its 
interaction with other galactic components (e.g., Wang et al. 2001, 2003;
Strickland et al. 2004). The arc-second resolution of these
observations, in particular, allows for 
a clean separation of point-like sources from the diffuse emission, 
a step critical for reliably determining both the content and physical 
condition of diffuse hot gas. 
Here I review key results from the observations
and discuss their implications.

\section{Summary of Existing {\sl Chandra} Observations}

\begin{table}[!ht]
\caption{A Sample of Nearby Normal Disk Galaxies Observed with {\sl Chandra}}
\smallskip
{\scriptsize
\begin{center}
\begin{tabular}{lccccccccc}
\tableline
\noalign{\smallskip}
Galaxy & Hubble & $D$ & Incl. &$L_{IR}$ & $L_{60}/L_{100}$& K$_{tot}$ & Env. & Exp. &$L_x$\\
Name   & Type & (Mpc) & (deg)&($10^{10} L_\odot$)& && & (ks)&($10^{39} L_\odot$)\\
\noalign{\smallskip}
\tableline
\noalign{\smallskip}
N4244 & Sd     & 3.6 & 85 &0.02 &0.26 &7.72& I & 60&$< 0.03$ \\
N4631 & Sd     & 7.5 & 85 &1.74 &0.40 &6.47& C & 60&2 \\
N3556 & Scd    & 14  & 80 &3.70 &0.42 &7.04& I & 60&3 \\
N3877 & Sc     & 17  & 76 &3.39 &0.33 &7.75& C &122&1 \\
N5775 & Sc     & 25  & 86 &5.25 &0.46 &7.76& C & 46&9 \\
N4565 & Sb     & 13  & 87 &0.77 &0.22 &6.06& C & 60&0.3 \\
N4594 & Sa     & 8.9 & 84 &0.20 &0.24 &4.96& C & 19&3 \\
\noalign{\smallskip}
\tableline
\end{tabular}
\end{center}
K$_{tot}$  - K band magnitude from the 2MASS survey;\par
Env. -- galaxy environment: I - isolated; C - with galaxy companions;\par
Exp. - {\sl Chandra} ACIS-S exposure time;\par
$L_x$ - Extra-planar diffuse 0.2-2 keV luminosity.
}
\end{table}

Table 1 lists a sample of nearby normal edge-on
galaxies observed with the {\sl Chandra} ACIS-S.
They span a broad range of morphological types and 
star forming properties as judged by the 60$\micron$ to 100$\micron$ intensity ratios
($L_{60}/L_{100}$; Table 1). 
But none of the sample galaxies is dominated by nuclear starburst or AGN 
activities. Both 
N4631 and N5775 are clearly disturbed by the interaction with their
companions. Others seem to be rather isolated. 
N4244, an extremely quite and intrinsically low-surface
brightness galaxy, presents little sign of diffuse X-ray emission (Strickland
et al. 2004). The remaining six galaxies
all show  strong evidence for large-scale extra-planar diffuse X-ray 
emission (Fig. 1).
Quantitative results have been reported only for N4631 and N3556 
 (Wang et al. 2001, 2003).
The analysis of the other galaxies is still ongoing, although crude estimates
of the extra-planar diffuse X-ray luminosities are included in Table 1. 

\begin{figure}[!bth]
\begin{center}
\hbox{
\psfig{figure=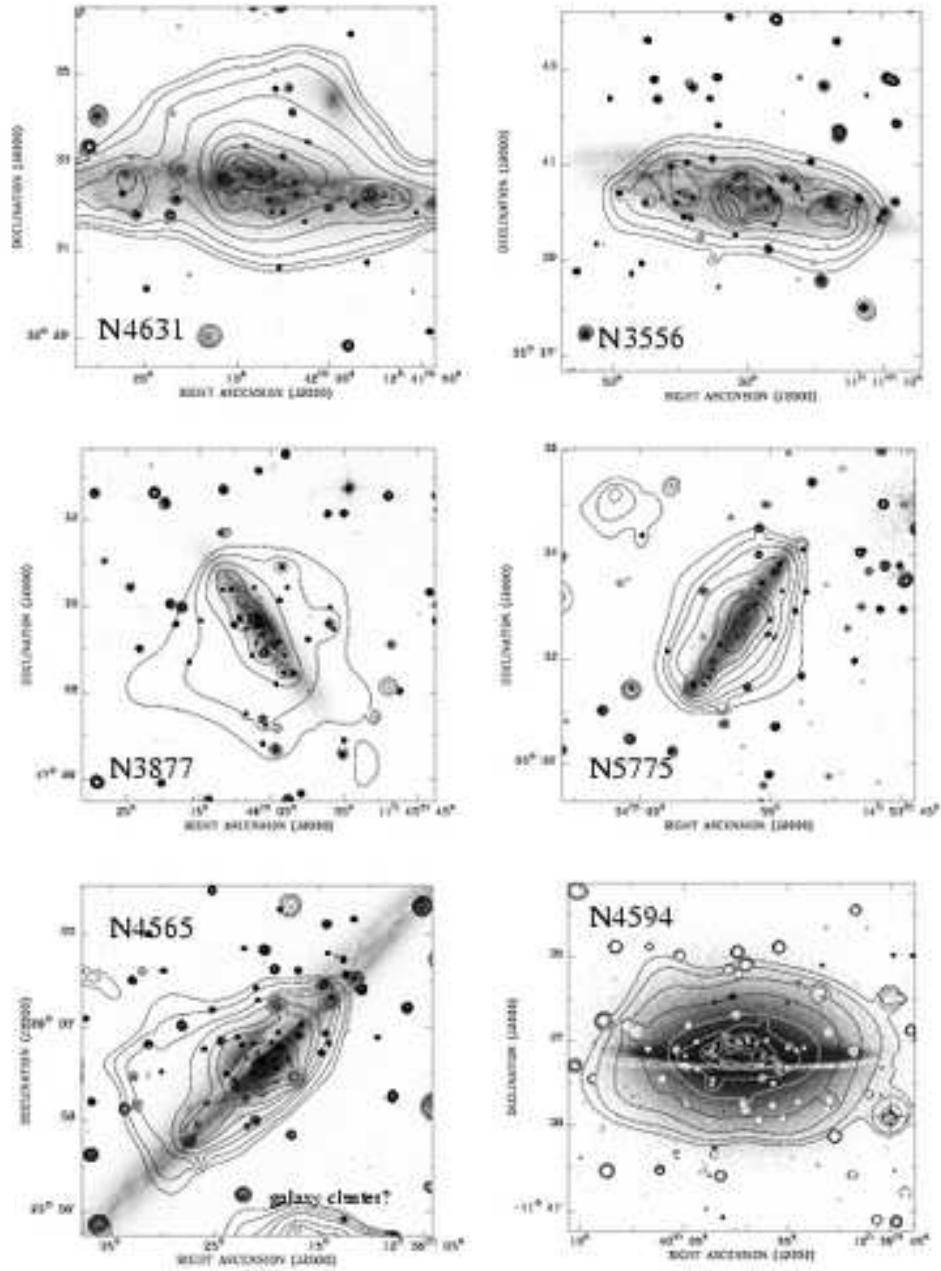,height=7.truein,angle=0.0,clip=}
}
\end{center}
\caption{(a) {\sl Chandra} ACIS-S intensity contours of nearby edge-on 
galaxies, overlaid on their optical B-band images. 
}
\end{figure}

\begin{figure}[!bth]
\begin{center}
\hbox{
\psfig{figure=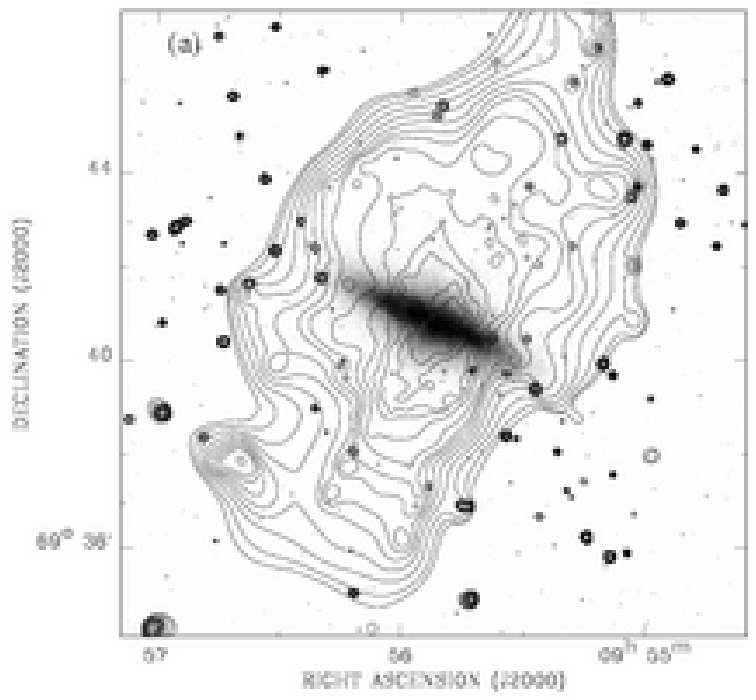,height=2.5truein,angle=0.0,clip=}
\psfig{figure=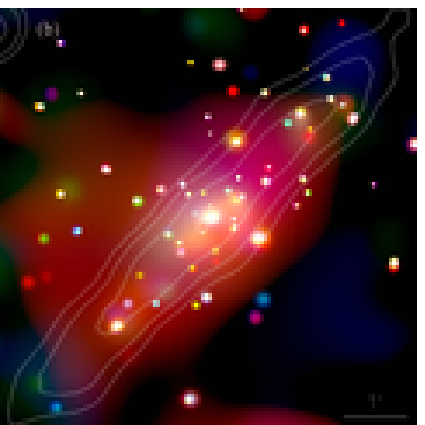,height=2.3truein,angle=0.0,clip=}
}
\end{center}
\caption{(a) {\sl Chandra} ACIS-I intensity contours of 
the nuclear starburst galaxy M82, overlaid on an
optical B-band image. The X-ray image is a mosaic of several existing 
observations;  CCD readout streaks have been subtracted.
(b) VLA 1.4 GHz continuum intensity contours overlaid on
the tri-color {\sl Chandra} ACIS-S image of N4565. The colors represent 
the intensities in the three energy bands: 0.3-0.7 keV (red),
0.7-1.5 keV (green), and 1.5-7 keV (blue). 
}
\end{figure}

It is interesting to compare the diffuse X-ray morphology of the normal
galaxies with that of nuclear starburst galaxies (e.g., Fig. 2a; see also 
Strickland et al. 2004).
The late-type (Sd-Sc) galaxies tend to show diffuse X-ray emission on scales
comparable to, or even larger than, the optical sizes.
The distribution of the galaxy-wide diffuse X-ray emission is 
relatively smooth and is more extended than H$_{\alpha}$-emitting
materials, which indicates that the bulk of the X-ray 
emission arises in galactic coronae --- hot gas confined around the 
host galaxies. Interestingly, the nuclear
starburst galaxy M82 (Fig. 2a) shows a simiarly 
large-scale diffuse X-ray-emitting halo, in addition to
an apparent bipolar outflow from the galactic central region
(Strickland et al. 2004). The X-ray morphology of the halo is 
more vertically elongated than those in Fig. 1. This elongation 
is likely caused by the large momentum of the nuclear outflow. 

In regions close to the galactic disks of both normal and nuclear starburst
galaxies, 
the X-ray emission appears to be rather filamentary and
shows a strong correlation with extra-planar  H$_\alpha$-emitting clouds, 
which are probably pushed out from the galactic disk
(Wang et al. 2001, 2003; Strickland et al. 2004 and references therein). 
Much of the X-ray emission may arise from freshly shocked cool gas clouds 
of small filling factor by fast-moving, low density outflows. 

The diffuse X-ray emission
seems to be more concentrated toward the inner regions of the early-type 
(Sb-Sa) galaxies. This concentration may be partly  due to the fact that
both N4565 and N4594 are very inactive in star formation. The gas heating
by Type Ia SNe from the old stellar population becomes more 
important than core-collapsed SNe. 

The diffuse X-ray spectrum can typically be
characterized with a thermal plasma of a few times 
$10^6$ K. The metal abundances appear to be enhanced in the diffuse hot gas:
O-like elements in late-type spirals and Fe-like elements in early-type
spirals (Fig. 3). But in general, both
the ionization and chemical states of the X-ray-emitting plasma are not
well constrained. The estimate of the absolute metal abundances, for example,
depends strongly on the
assumed plasma emission model, e.g., the temperature distribution.
The X-ray luminosity estimate of a low-surface brightness corona
is sensitive to the exact background subtraction. Nevertheless, 
we find that the luminosity appears to be proportional to the star formation 
rate and to the stellar mass, as traced by the far-IR and K-band luminosities 
of the galaxies, respectively. But the diffuse X-ray luminosity typically accounts for 
less than a few percent of the expected SNe energy input in each 
galaxy.

\begin{figure}[bth]
\begin{center}
\hbox{
\psfig{figure=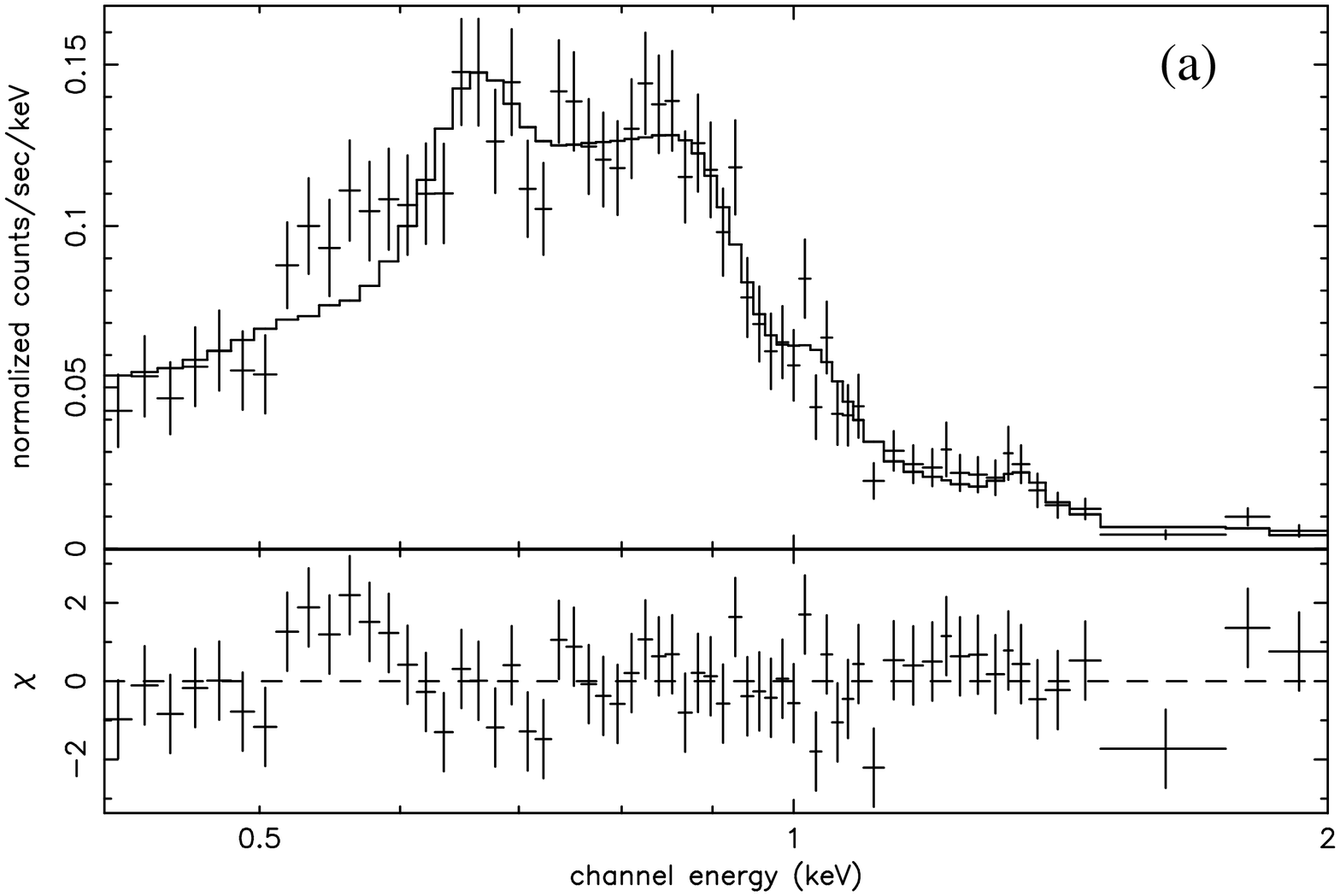,height=1.7in,angle=0,clip=}
\psfig{figure=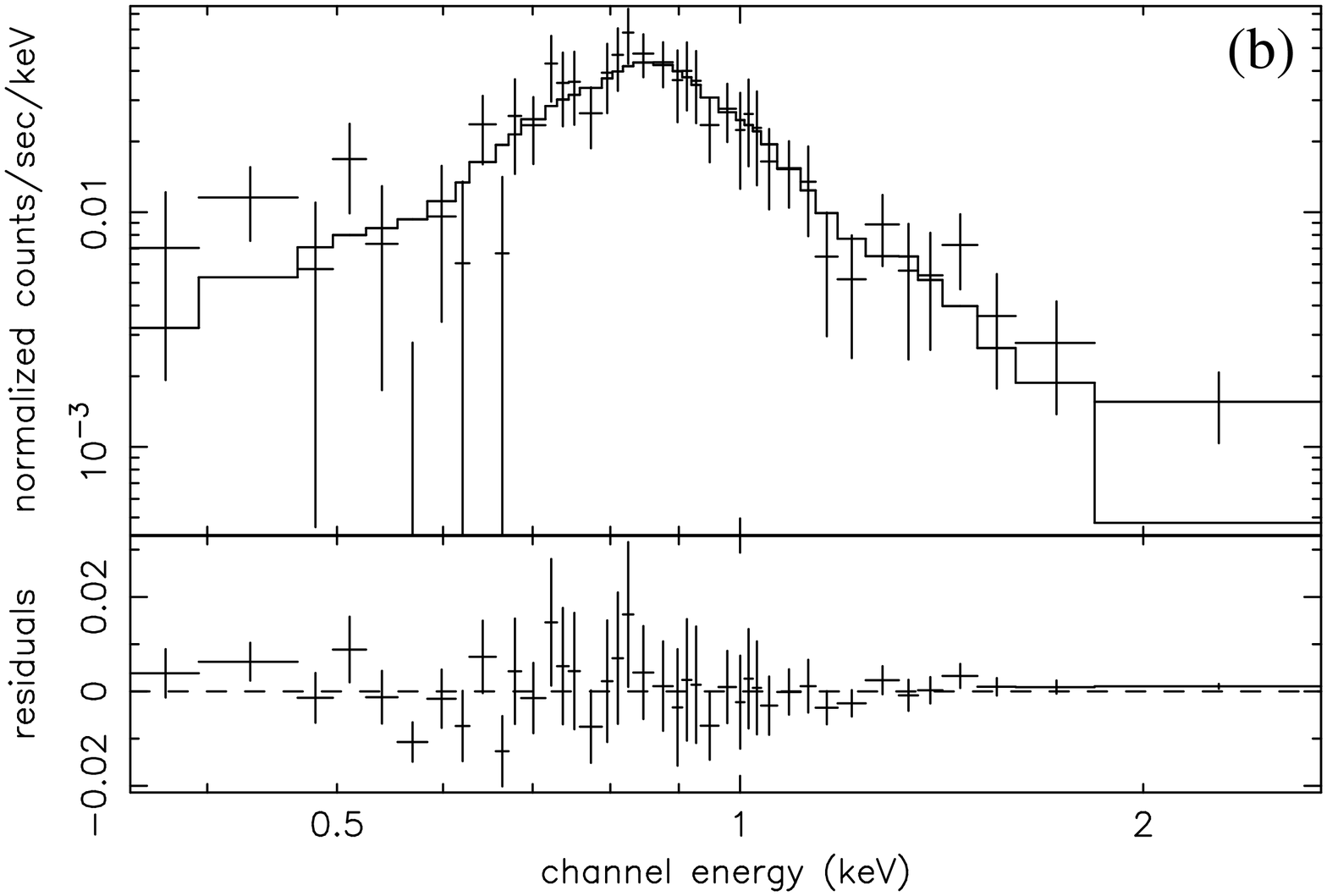,height=1.7in,angle=0, clip=}
}
\vspace{-1.0cm}
\end{center}
\caption{
(a) The {\sl Chandra} ACIS-S spectra of the diffuse emission from the central 
corona region of N4631. The histogram represents the 
best-fit thermal plasma plus a power law
(photon index = 1.5; accounting for $\sim 10\%$ counts), 
constrained with a joint fit to 
an integrated spectrum of discrete sources. The best-fit temperature is 
$\sim 0.4$ keV, while 
the metal abundances are 0.8 and 0.2 solar for the O-like and Fe-like 
elements, 
respectively. (b) An ACIS-S spectrum of the extra-planar diffuse emission
from N4594  and the best-fit galactic bulge-outflow model (histogram).
Notice the strong and extended Fe-L bump at $\sim 0.8 $ keV.
}
\end{figure}

\section{Implications}


\subsection{Mechanical Energy Balance}

The mechanical energy balance in the ISM represents one of the basic and 
unsolved issues in galaxy evolution. A large fraction of SN 
blastwave energy is expected to be in the X-ray-emitting gas. But the
energy is apparently not radiated in X-ray. Where does the SN energy go?

For late-type spirals with rich cool gas, the ``missing'' energy is 
perhaps radiated in a lower energy band
than the X-ray, such as the UV or IR.  
In the N4631 corona, for example, a considerable fraction of the 
energy may be accounted for by  the emission of the O{\small VI} 
1032/1038\,\AA\ doublet (Otte et al. 2003). At temperatures of a few 
times $10^5$\,K, where gas of near-solar metallicity cools most efficiently, 
the O{\small VI} doublet is the dominant coolant.  With
some reasonable extrapolations, assuming an overall morphology and 
an extinction correction in these regions, the O{\small VI}-inferred 
cooling can account for a substantial fraction of the SN energy input. 
However, 
the nature of O{\small VI}-bearing gas remains uncertain. It may represent 
large-scale diffuse cooling gas or may only reside at the interfaces between 
diffuse hot gas and cool clouds of small filling factor. 
The existing X-ray CCD data (e.g., Fig. 3) have too limited spectral 
information content to 
uniquely determine the thermal state of hot gas, even under constrained 
assumptions about its ionization and chemical states. 
The estimate of the total radiative cooling rate, in particular, depends 
sensitively on the extrapolation of the assumed spectral shape. The link
between the gas components observed in far-UV and X-ray is still missing.

Furthermore, N4631 may not be a typical normal galaxy. It is strongly
interacting with its companions, as evidenced by both the presence of
H{\small I} tidal arms and the enhanced star formation in the galactic disk.
A more pristine environment for examining the disk/halo
interaction can be found in the relatively
isolated edge-on Sc galaxy such as N3556, around which the extra-planar 
diffuse X-ray-emitting gas has also been detected (Wang et al. 2003; Fig. 1b).
We have obtained 120 ksec {\sl FUSE} observing time in the current cycle to 
further our investigation of the role that O{\small VI}-bearing gas 
might play in balancing the SN energy input.

A significant fraction of the SN energy may also be
carried away by cosmic rays diffusing out of galaxies. Such outflows
can be traced by extra-planar radio continuum emission and may be
accompanied by outflows from galactic disks. Indeed, there is an overall 
morphological similarity between the
extra-planar radio continuum and diffuse soft X-ray emission
from late-type disk galaxies (Wang et al. 2001, 2003).
But the similarity is absent in early-type disk galaxies.
Take the Sb galaxy N4565 as an example (Fig. 2b). There is no evidence for
significant extra-planar radio continuum emission, whereas
the diffuse soft X-ray emission extends as far as $\sim 15$ kpc vertically.
Therefore, the cosmic ray diffusion cannot be important. The missing energy
problem remains.
The total luminosity of the diffuse X-ray emission from the galaxy is 
$\sim 2 \times 10^{39} {\rm~erg~s^{-1}}$, only about 3\% of 
the expected input from Type Ia SNe alone. A similar fraction is found for
N4595. It is important to note that Type Ia SNe 
occur primarily in the
galactic bulges or halos, in which there is too little cool gas to hide or 
convert the energy.

Most likely, galactic bulge outflows driven primarily by Type Ia SNe 
are responsible for much of the mass and energy balance
in early-type disk galaxies, similar to the galactic winds
proposed originally for elliptical galaxies (Mathews \& Baker 1971; 
Bregman 1980b). 
We are quantitatively testing this idea, based on a simple spherically symmetric outflow model (Li \& Wang 2004). 
We assume that the mass-loss and energy input rates are proportional to
the near-IR star light and account for the galactic gravitational potential 
 in determining the gas dynamics. We can now construct XSPEC 
models
for both the surface brightness profile (in any selected energy band) and 
the projected spectrum (in any selected region), which can be analyzed in
the same fashion. Preliminary fits to the X-ray data for N4595
show reasonably good fits (e.g., Fig. 3b). The fitted total mass and energy 
input rates are consistent with the values expected from the optical or 
near-IR luminosities of the galactic bulge. Most interestingly, the fits
indicate that the outflow is particularly rich in iron; the abundance is
$\sim 4 \times$ solar. However, there are some considerable discrepancies 
between the model and the data.
The observed surface brightness seems to be systematically higher than 
predicted in outer regions of the galactic halo, which might
indicate the limitation of the 1-D galactic outflow model and/or 
the importance of the interaction between the outflow and the accretion
from the intergalactic medium (IGM).
   
\subsection{ISM-IGM Connection Around Disk Galaxies}

Ultimately, a comprehensive understanding of the extra-planar hot gas
has to be put in the
context of galaxy evolution. It is widely believed that 
disk galaxies are still accreting from the IGM, replenishing cold gas that is 
consumed by star formation (e.g., Toft et al. 2002 and references therein).
Nevertheless, the gas around a massive galaxy may be heated to X-ray-emitting 
temperatures by an accretion shock and gravitational compression. 
But, the existing simulations do not adequately account for the energy feedback 
from the galaxies, especially the heating due to Type Ia SNe and the
associated outflows discussed above.
The predicted temperatures seem to be consistent with those inferred
from the observations. But, there is a significant discrepancy between the
observed extra-planar diffuse X-ray luminosities and the predictions 
from the simulations (Fig. 4). 
The observed X-ray luminosity of N4594,
for example, is about a factor of $\sim 30$ lower than the predicted value.

\begin{figure} 
\begin{center}
\psfig{figure=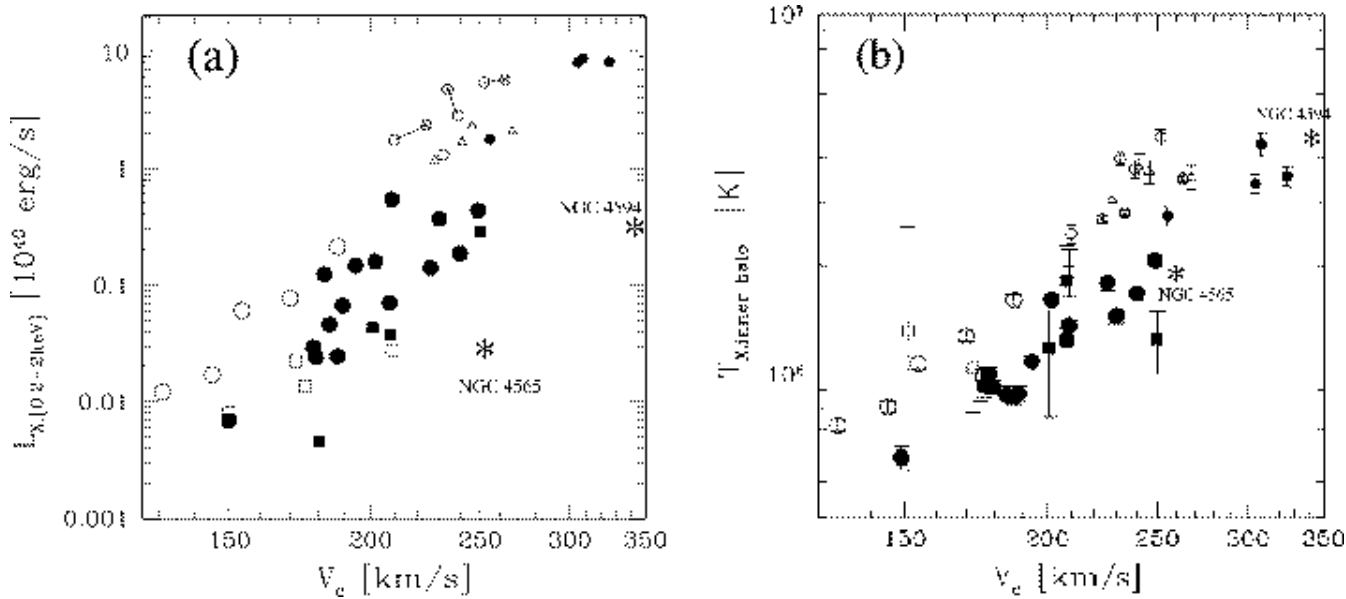,height=3.1in,angle=270, clip=}
\end{center}
\vspace{-1.0cm}

\caption{Observed diffuse X-ray properties of N4565 and N4594
(marked as {\sl stars}),
compared with the predictions from numerical simulations of disk galaxy 
formation (Toft et al. 2002):
0.2--2~keV band luminosity  (a) and average inner halo hot 
gas temperature  (b) versus galaxy characteristic 
circular speed. Various symbols represent simulations assuming different
cosmologies. The large filled symbols, in particular, are from simulations 
with $(\Omega_\Lambda,\Omega_M) = (0.7, 0.3)$ cosmology with the baryon
fraction $f_b = 0.1$: Circles correspond to primordial abundance, whereas
squares to $Z = 1/3 Z_\odot$.
}
\end{figure}

The under-luminosity of the accretion may be due 
to the effect of the outflows from the galactic bulges, which tends to
reduce the emission measure in the inner regions of the galaxies.
The asymmetric morphology of diffuse extra-planar X-ray emission around N4565
may further suggest a complicated galactic outflow/accretion interaction. One may
expect that the bulge outflow will eventually be stopped by the ambient medium. 
The reverse-shocked outflow materials may then accumulate around galaxies, responsible
for much of the large-scale extra-planar diffuse soft X-ray emission. 
But if a galaxy has a motion relative to the ambient IGM, the outflow materials
may be swept into a trail by the
ram-pressure, which might be amplified by the gravitational focusing.
In the case of N4565, the motion may be toward the northeast. The outflow in
this direction may be confined in region close to the disk, enhancing
the diffuse X-ray emission. But simulations need to be done to see whether
or not such a scenario may explain the observed morphological and
spectral characteristics of the emission.

\section{Future Prospect}

Our ongoing systematic analysis of the existing {\sl Chandra} data,
together with the dedicated modeling, 
will give a more quantitative test of the various 
scenarios mentioned above. But a real breakthrough probably 
requires a new generation of observing tools. In the near future,
{\sl Astro-E2} high spectral resolution observations of edge-on galaxies can provide unique
spectroscopic diagnostics of the thermal, chemical, and ionization 
states of the extra-planar hot gas, which will then allow 
 a definitive test of various assumptions made in the analysis 
of X-ray CCD spectra and the role of extra-planar diffuse hot gas 
in both balancing the mechanical energy input from SNe and 
re-distributing chemical enriched materials in galaxies.

Another diagnostic tool that {\sl Astro-E2} may offer is 
the measurement of emission line
broadening. According to the ground calibration data,
the width of a strong emission line can be determined to $\la 0.5$ eV ($\sim 150 
{\rm~km~s^{-1}}$). This resolving power 
results mainly from the fact that the 
line response profile of the instrument 
is almost perfectly Gaussian to 1 part in $10^4$. 
This capability raises the possibility to determine the velocity dispersion
of extra-planar diffuse hot gas. Indeed, the supposedly
much cooler OVI-bearing gas, as seen in the {\sl FUSE} spectra of
N4631, shows a substantial broadening with 
a $FWHM \sim 200 {\rm~km~s^{-1}}$, corresponding to a thermal 
temperature of $\sim 10^7$ K (Otte et al. 2003). One may expect that
the X-ray-emitting gas should have even broader lines. With a 
reasonable counting statistics, a line broadening of $\sim 300 {\rm~km~s^{-1}}$
could easily be resolved. These new capabilities are essential to 
further the study of the heating, transferring, and cooling of extra-planar
diffuse hot gas in galaxies.

I thank my collaborators for their contributions to the project described
above and supported by NASA through the grant GO1-2084A.

\section{References}
Bregman, J. N. 1980a, ApJ, 236, 577
\\ Bregman, J. N. 1980b, ApJ, 237, 280
\\ Li, Z. Y., \& Wang, Q. D. 2004, in preparation
\\ Mathews, W. G., \& Baker, J. C. 1971, ApJ, 170, 241
\\ McKee, C. F., \& Ostriker, J. P. 1977, ApJ, 218, 148
\\ Norman, C. A., \& Ikeuchi, S. 1989, ApJ, 345, 372
\\ Otte, B., et al. 2003, ApJ, 591, 821
\\ Spitzer 1956, ApJ, 124, 20
\\ Strickland, D. K., et al.  2004, ApJ, 606, 829
\\ Toft, S., et al. 2002, MNRAS, 335, 799
\\ Wang, Q. D., et al. 2001, ApJL, 555, 99
\\ Wang, Q. D., Chaves, T., \& Irwin, J. A., 2003, ApJ, 598, 969
\\

\end{document}